\documentclass[letterpaper,12pt, abstracton]{scrartcl}

\usepackage{marvosym}
\usepackage[T1]{fontenc}
\usepackage{color,hyperref}
\usepackage{enumitem}
\usepackage{amssymb}
\usepackage{amsmath}
\usepackage{wasysym}
\usepackage{pifont}
\usepackage{makecell}
\usepackage{subcaption}
\usepackage{booktabs}
\usepackage{Sweave}
\newcommand{\ra}[1]{\renewcommand{\arraystretch}{#1}}

\date{}

\begin{document}

\title{Statistical look at reasons of involvement in wars}
\author{Igor Mackarov \\ \href{mailto:Mackarov@gmx.net}{{\small Mackarov@gmx.net}}}
\Sconcordance{concordance:Mackarov.tex:Mackarov.rnw:%
1 17 1 1 0 27 1 1 8 7 0 1 5 6 0 1 2 3 1 1 2 4 0 1 2 2 1 1 2 1 0 1 1 12 %
0 1 2 4 1 1 7 6 0 2 1 1 4 2 0 1 2 1 0 1 4 2 0 3 1 1 3 1 0 1 2 4 0 1 2 3 %
1 1 6 10 0 1 5 14 0 1 2 3 1 1 2 4 0 1 2 11 1 1 5 3 0 1 3 1 5 3 0 1 9 7 %
0 1 2 1 3 1 0 2 1 12 0 1 2 6 1 1 4 13 0 1 2 13 1 1 2 1 0 1 2 50 0 1 2 1 %
3 1 0 1 1 1 3 5 0 1 3 24 1}

\maketitle
\begin{abstract}
The Correlates of War project scrupulously collects the information about disputes between the countries over a long historical period together with other data relevant to the character and reasons of international conflicts. Using methods of modern Data Science implemented in the R statistical software, we investigate the datasets from the project. We study political, economic, and religious factors with respect to the emergence of conflicts and wars between the countries. The results obtained lead to certain conclusions about variances and causalities between the factors considered. Some unpredictable features are presented.

\textbf{Keywords:}  \textit{Correlates of War, Variance, Factorial ANOVA, Normality, R.}

\end{abstract}

\section{Introduction}

Here we study datasets from the  \href{http://www.correlatesofwar.org/}{Correlates of War} (COW) project which provides accurate and reliable quantitative data in international relations, in particular, in the disputes between countries over a long historical period together with other information relevant to the character and reasons of the conflicts.

 \textit{Diplomatic, economic, and religious} aspects are among the factors affecting emergence of international tensions. In this paper we have investigated them using statistical methods implemented in the \textit{R} data analysis software.

\section {Distribution of countries over the number of their wars}
\label{sec:distrib}

\begin{figure}[h]
  \centering
  \includegraphics[width=.7\textwidth]{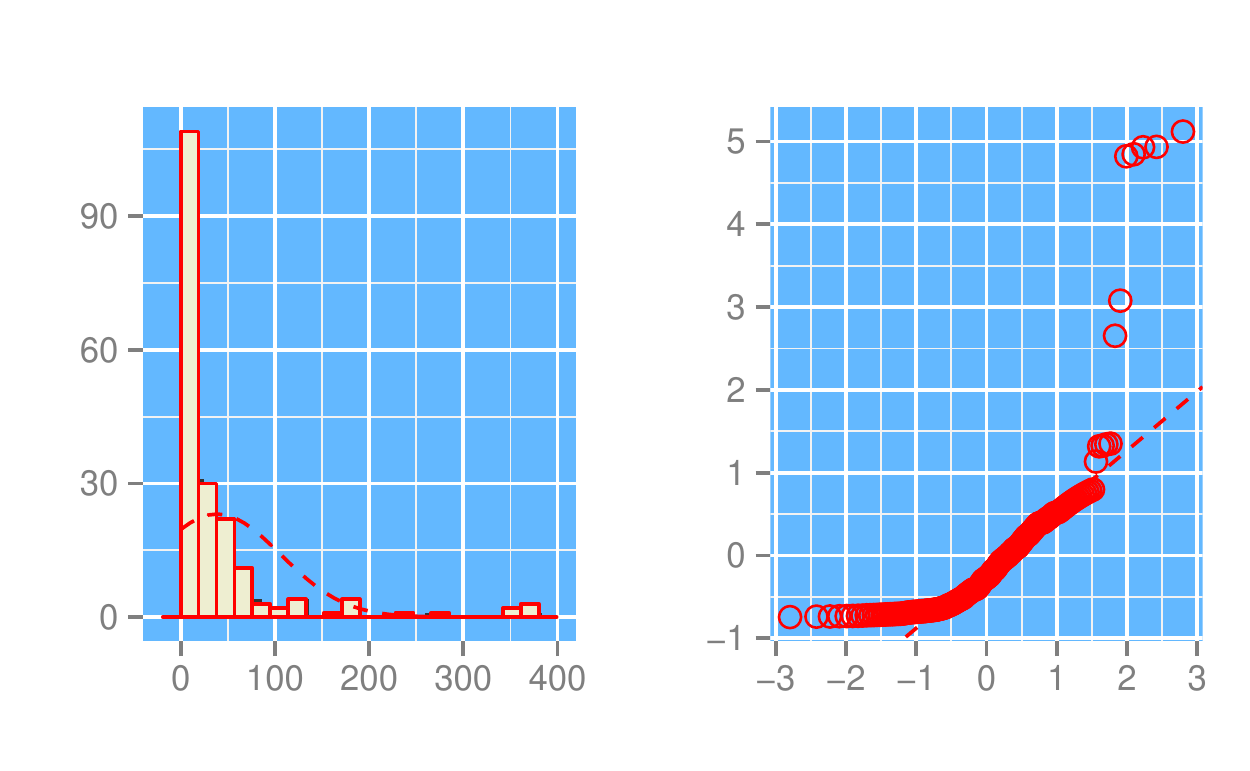}\\
  \caption{Histogram -- distribution of countries (on their codes) by the numbers of wars they have been involved in over the period 1816--2010 together with Q--Q plot for this distribution}\label{histo}
\end{figure}

 Before looking at the above-depicted factors, let us range the countries on their ``belligerence'', i. e., the number of wars they have been involved in throughout the period 1816--2010 \cite{COW1}. The distribution turns out \textit{normal} at whatever level of statistical significance as suggested by the results of four basic normality tests executed on\textit{ R}:

\begin{scriptsize}
\begin{Schunk}
\begin{Sinput}
> shapiro.test(no_of_wars)
\end{Sinput}
\begin{Soutput}
	Shapiro-Wilk normality test

data:  no_of_wars
W = 0.55004, p-value < 2.2e-16
\end{Soutput}
\begin{Sinput}
> library(nortest)
> ad.test(no_of_wars)
\end{Sinput}
\begin{Soutput}
	Anderson-Darling normality test

data:  no_of_wars
A = 26.233, p-value < 2.2e-16
\end{Soutput}
\begin{Sinput}
> cvm.test(no_of_wars)
\end{Sinput}
\begin{Soutput}
	Cramer-von Mises normality test

data:  no_of_wars
W = 4.9752, p-value = 7.37e-10

Warning message:
In cvm.test(no_of_wars) :
  p-value is smaller than 7.37e-10, cannot be computed more accurately
\end{Soutput}
\begin{Sinput}
> lillie.test(no_of_wars)
\end{Sinput}
\begin{Soutput}
Lilliefors (Kolmogorov-Smirnov) normality test

data:  no_of_wars
D = 0.28451, p-value < 2.2e-16
\end{Soutput}
\end{Schunk}
\end{scriptsize}

 Yet, as shown by the Q--Q plot at Fig. \ref{histo}, there is a good deal of ``outliers'' which turned out involved in the largest number of wars. The top 12  most ``war-full'' countries are thus \textbf{\textit{USA, Russia, UK, China, Germany, France, Japan, Turkey, Iran, Italy, Israel, and Iraq.}} They swerve from the normal distribution of the rest of the countries. This is quit a natural list since those are the countries with complicated history (participants of both world wars or parts of regions with traditionally high military activity). 
 
 Evidently, these countries are of the most interest regarding the studies of the wars reasons. Later on we will then focus on the countries from this list.

\section{Examining principal factors of wars}

\subsection{Political relations at times of war and peace}

 \paragraph[short]{Variables.}To estimate the value of diplomacy in disputes between a pair of countries, two factor variables were considered: \textit{Diplomatic representation level (\textit{DR})} of side 2 at side 1 and vice versa, and \textit{Any diplomatic exchange} (\textit{DE})  between sides 1 and 2 \cite{COW2, Jones}. \textit{DR} can have following values:

 \begin{itemize}[label=$\checkmark$]
 \item 0=no evidence of diplomatic exchange
 \item 1=charg\'e d'affaires
 \item 2=minister
 \item 3=ambassador
 \item 9=other
 \end{itemize}

 \paragraph[short]{Measurement scales, missing values treatment.}

 The last level of \textit{DR} looks like an empty value and needs to be handled accordingly. Level 9 means ``interest sections, interests served by another country, address only'' etc. \cite{COW2}, i.e., kind of ``minor'' representation.  It seemed thus reasonable  to rearrange the levels in such a way that level 9 took the place between levels 0 and 1. Since then the variable was treated as \textit{ratio} rather than \textit{nominal}.
\\~

The  \textit{DE} variable may equal:
\begin{itemize}[label=$\checkmark$]
 \item0=neither side was represented in the other side
 \item1=at least one side was represented in the other side
\end{itemize}
It was also treated as a \textit{ratio} variable.

 We thus consider the variance of  \textit{DR} and \textit{DE}  with respect to two categorical variables, viz.\textbf{ \textit{a country}}  (1 or 2, the countries -- a dispute opponents), and \textbf{\textit{a state}} (war or peace).

Perform a two factor analysis of variance (ANOVA) \cite{Nutshell} to investigate the relations of interest.

 Instead of investigating the countries pairwise, we chose a few of them and considered each versus \textit{all the countries} it ever had disputes with in 1817--2005. Such a method gave much larger and homogeneous data samples compared to the pairwise approach, which better satisfied the ANOVA assumptions (see the next subsection).

 \paragraph[short]{Test results.} Table \ref{tab:dipl} shows the results of the two factor ANOVA test on the relation between \textit{DR} and the two chosen factors. These factors are: a certain country on the one hand and all its opponents on the other hand, and a war/peace state.  Also shown is the significance of correlations with the combination of the first and second factors. Again, the one factor correlation is presented of the \textit{DE} variable (common for both sides) with the war/peace state.
  \begin{table}
  \centering
  \caption{Results of tests of dependencies between diplomatic relations of a few selected countries with their opponents at times of peace and war. As per the standard notation \cite{Lantz}, symbols `***', `**', `*', `.' to the right of the numbers mean statistical significance at levels 0.1\%, 1\%, 5\%, 10\%, respectively}
  \label{tab:one}
  \ra{1.3}
  \begin{tabular}{@{}rrrrcrrrcrrr@{}}\toprule
  & \multicolumn{3}{c}{Representation levels on both sides (\textit{DR})} & \phantom{abc}& \multicolumn{3}{c}{\ \ \ \textit{DE}\ \ \ }\\
  \cmidrule{2-4} \cmidrule{6-8}
  &country - its opponents& war-peace & interaction &&  war-peace  \\ \midrule
  \textbf{Top-12 list} \\ \textbf{countries:}\\
  $USA$ & 1.83e-07 *** & 2.65e-08 *** & 0.08. &&2e-16***  \\
  $Russia$ & 5.38e-06***& 2.2e-16***&0.71&&1.08e-09***\\
  $UK$ &7.62e-11***&0.061.&0.52&& 0.20\\
  $China$ &0.20&0.70& 0.94&& 0.23\\
  $Germany$ & 2.2e-16***& 0.0015**& 0.72&& 0.00087***\\
  \textbf{Others:}\\
  $North Korea$ &0.73&0.13&0.90&&0.073.\\
  $Libya$ &0.92&0.0061**&0.75&&0.56\\
  $Somalia$ &9.3e-05***&0.42& 0.98&&0.049*\\
  \bottomrule
 \label{tab:dipl}
  \end{tabular}
  \end{table}

 Evidently, in most of the cases there indeed is a statistically significant relation between diplomatic activities on any side and the state (war/peace) the opponents are in. Such a significance was observed for a much larger number of countries examined.

\paragraph{Exception.} Note, however, a couple of rare violations of this trend: absence of statistical significance in diplomacy of \textit{China}  and  \textit{North Korea}, even though both have pretty long peaceful and tense history. What is a common between these countries? In particular, a strong political isolation at times of communist past and present. Perhaps, this is the reason of low diplomatic response to the war/peace state in these cases.

\subsubsection{Meeting the ANOVA assumptions}
\label{assunptions}

For a correct use of ANOVA following assumptions are known to be satisfied:

\newenvironment{MYitemize}{%
\renewcommand{\labelitemi}{$\leadsto$}%
\begin{itemize}}{\end{itemize}}
\begin{MYitemize}
\item \textit{normality}
\item \textit{homogeneity of variance}
\item \textit{independence of observations}
\end{MYitemize}

In that regard, all the tests were accompanied by checking the normality of distributions withing the samples, as well as their covariances. In virtually all the cases the normality was great, like in the case ``China--opponents'':
\begin{scriptsize}
\begin{Schunk}
\begin{Sinput}
> library(nortest)
> lillie.test(sample1)
\end{Sinput}
\begin{Soutput}
	Lilliefors (Kolmogorov-Smirnov) normality test

data:  sample1
D = 0.25952, p-value < 2.2e-16
\end{Soutput}
\begin{Sinput}
# ...
\end{Sinput}
\begin{Soutput}
data:  sample2
D = 0.24207, p-value < 2.2e-16
\end{Soutput}
\begin{Sinput}
# ...
\end{Sinput}
\begin{Soutput}
data:  sample3
D = 0.29846, p-value < 2.2e-16
\end{Soutput}
\begin{Sinput}
# ...
\end{Sinput}
\begin{Soutput}
data:  sample4
D = 0.24584, p-value < 2.2e-16
\end{Soutput}
\begin{Sinput}
# ...
\end{Sinput}
\begin{Soutput}
data:  sample_general
D = 0.49108, p-value < 2.2e-16
\end{Soutput}
\end{Schunk}
\end{scriptsize}
The variances of the four groups of   \textit{DR} values (corresponding to all the combinations of values ``side 1, side 2''---``war, peace'') were comparable like in the case involved:
\begin{scriptsize}
\begin{Schunk}
\begin{Soutput}
variances:
sample1= 1.33
sample2= 1.58
sample3= 1.12
sample4= 1.49
\end{Soutput}
\end{Schunk}
\end{scriptsize}
This evidently assures homogeneity of the samples variance.

As to the independence of observations, it is guaranteed the by standard scientific principles of the data collection on the COW project.

\subsubsection{All the countries.} Finally, we  performed a general test. For the \textit{two factor} ANOVA we also had  four samples like the ones mentioned in subsection \ref{assunptions}. They were such that sides 1, 2 involved the diplomatic activity data (\textit{DR}) for the years of disputes of \textit{any country} against \textit{all the countries} it ever was in the state of a dispute, as well as the data for the correspondent pairs of the countries at peaceful times.
\begin{scriptsize}
\begin{Schunk}
\begin{Sinput}
> anova( lm(dip_presence ~ countries*state))
\end{Sinput}
\begin{Soutput}
Analysis of Variance Table

Response: dip_presence
                                    Df Sum Sq Mean Sq F value Pr(>F)
countries                    1      0 0.00000   0.000 1.0000
state                        1      0 0.09898   0.063 0.8019
countries:state              1      0 0.00000   0.000 1.0000
Residuals               263880 414910 1.57234               
\end{Soutput}
\end{Schunk}
\end{scriptsize}
For the \textit{DE} variable, common for a pair of opponents, the results of the\textit{ one factor }test are as follows:
\begin{scriptsize}
\begin{Schunk}
\begin{Sinput}
anova( lm(dip_exchange ~ state))
\end{Sinput}
\begin{Soutput}
Analysis of Variance Table

Response: dip_exchange
                                    Df Sum Sq Mean Sq F value    Pr(>F)   
state                                1      3  3.4820  20.384 6.337e-06 ***     
Residuals                       263880  45075  0.1708                      
---
Signif. codes:  0 '***' 0.001 '**' 0.01 '*' 0.05 '.' 0.1 ' ' 1
\end{Soutput}
\end{Schunk}
\end{scriptsize}
\paragraph[short]{Analysis.} As per the test results, the general level of diplomatic activity of opponents (\textit{DE}) vs. the war/peace state is the only statistically significant relation. Apparently, this means that the different opponents in different years could have different structures of political activity with respect to each other but these structures are ``averaged'' over the large number of countries and years without yielding any characteristic pattern, common for all the countries.

What is indeed common for all the countries, it is the sensitivity of the joint level of their diplomatic activity to essential conditions of the interstate relations dictated by essential circumstances --- war or peace (\textit{DE} vs. war/peace).

\begin{table}[!htbp]
\centering
\caption{Results of tests on dependencies between \textit{Composite Index of National Capability} of a few selected countries vs. that of their opponents at  times of peace and war}
\label{tab:two}
\ra{1.3}
\begin{tabular}{@{}rrrrcrrrcrrr@{}}\toprule

& country - its opponents& war-peace & interaction of both factors  \\ \midrule
\textbf{Countries from} \\ \textbf{top-12 list:}\\
$USA$ & 2.2e-16*** &0.062.  & 0.017*   \\
$Russia$  &2.2e-16***&0.00117**&0.00104** \\
$UK$  &2.2e-16***&0.018*&0.45\\
$China$ &2.2e-16***&1.9e-13 ***&2.2e-16***\\
$Germany$ &2.2e-16***&0.010*&0.97\\
\textbf{Others:}\\
$North Korea$ &2.2e-16***&1.84e-07***&5.2e-07***\\
$Libya$ &2.0e-16***&0.077.&0.099. \\
$Somalia$ &2e-16***&0.0094***&0.011*  \\
\bottomrule
\end{tabular}
\label{tab:econom}
\end{table}

\subsubsection*{}

\subsection{Economic state at times of war and peace}
Among a great number of the countries' economic life characteristics provided by the ``economic'' datasets \cite{COW_material}, we chose the common aggregated parameter, \textit{Composite Index of National Capability (CINC)}.

\paragraph[short]{Non-available values.} A small number of NAs was found. Since \textit{CINC} may change essentially throughout the historical period, instead of traditional replacement of NAs by a median or mean value, it was found reasonable to use a valid \textit{CINC} \textit{of the nearest year} for the replacement.

\paragraph{Results of the tests.}   Table \ref{tab:econom} demonstrates the results of two factor ANOVA for the same set of countries as in the diplomatic case. We see that the statistical significance here is evident and even better then in Table \ref{tab:dipl}. This makes consider the economic factor most objective indicator of the quality of interstate relations.\\

\begin{figure}[t]
        \centering
        \begin{subfigure}[]{0.4\textwidth}
                \includegraphics[width=1\textwidth]{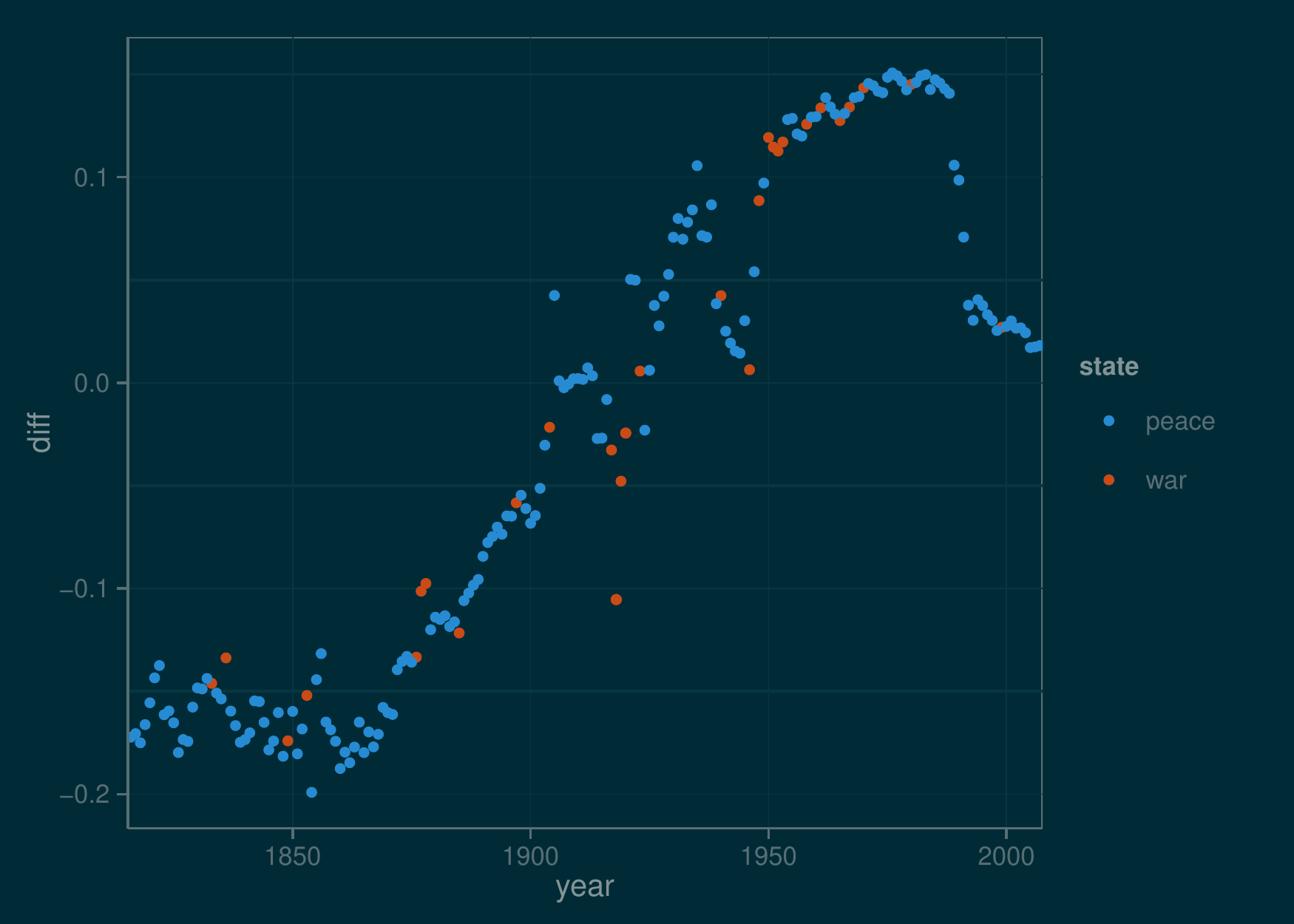}
                \caption{Russia---UK}
        \end{subfigure}
        \begin{subfigure}[]{0.4\textwidth}
                           \includegraphics[width=1\textwidth]{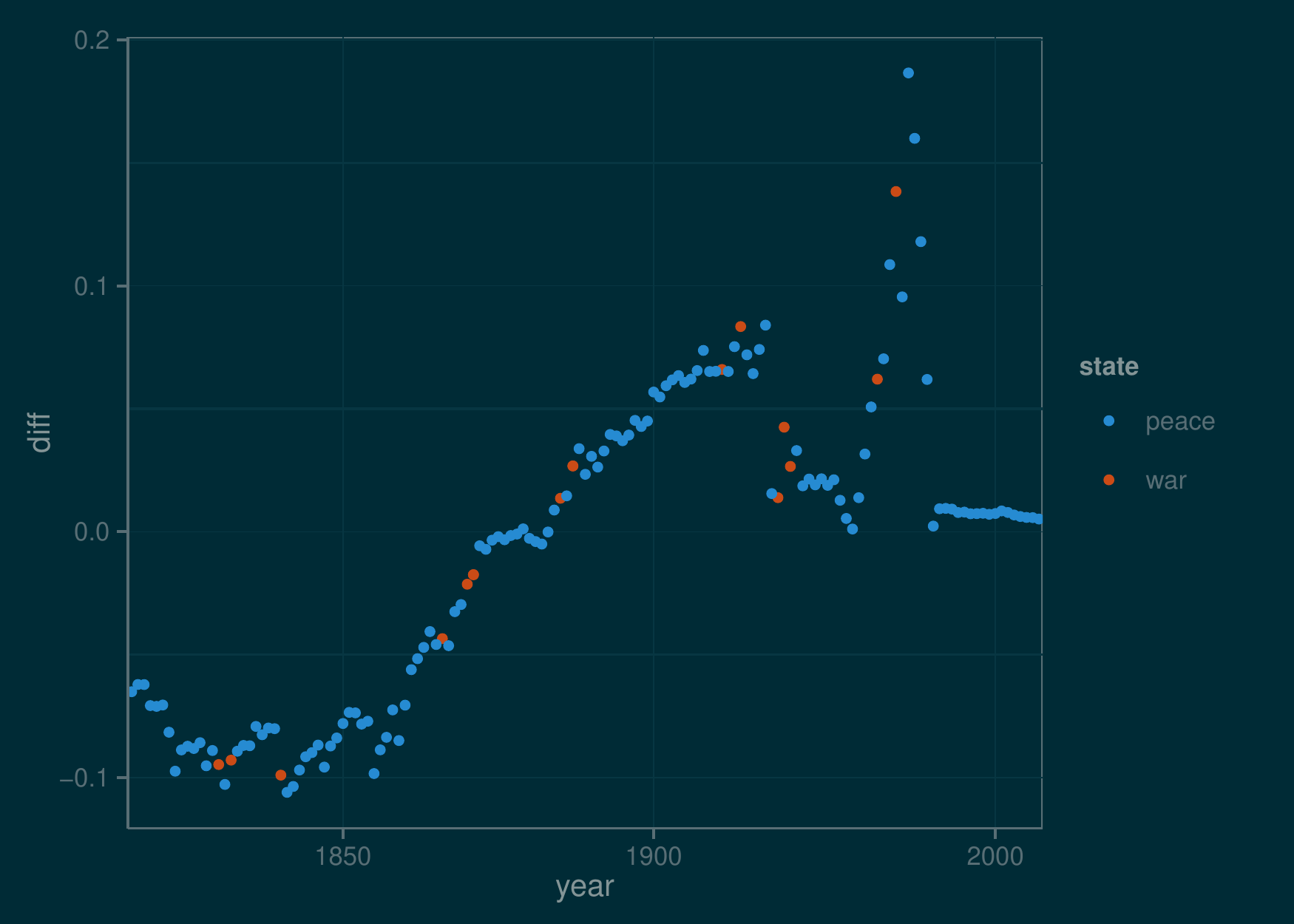}
                           \caption{Germany--France}
          \end{subfigure}
\\~
\\~

        \begin{subfigure}[]{0.4\textwidth}
                           \includegraphics[width=1\textwidth]{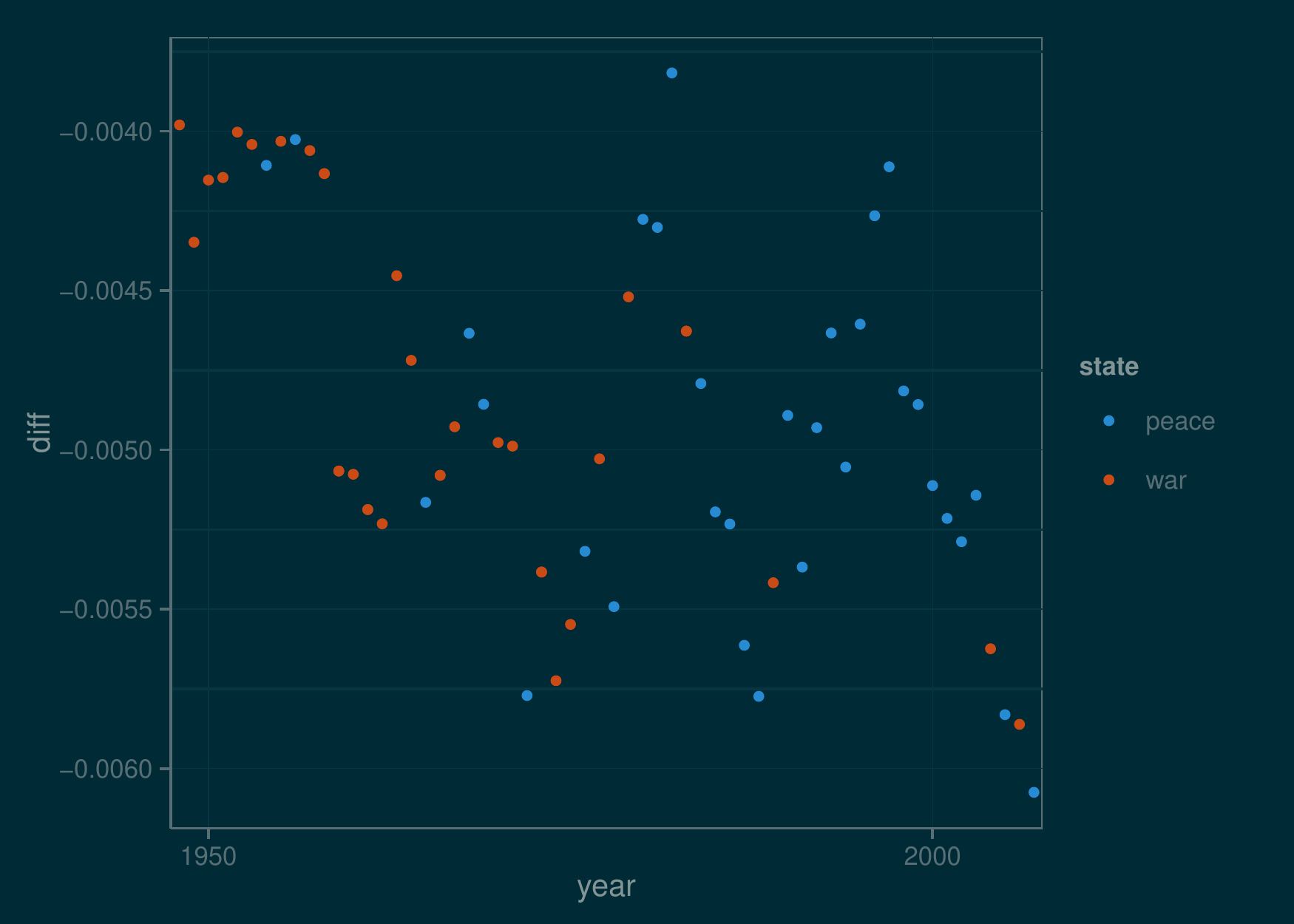}
                           \caption{Israel--Egypt}
       \end{subfigure}
      \begin{subfigure}[]{0.4\textwidth}
                                  \includegraphics[width=1\textwidth]{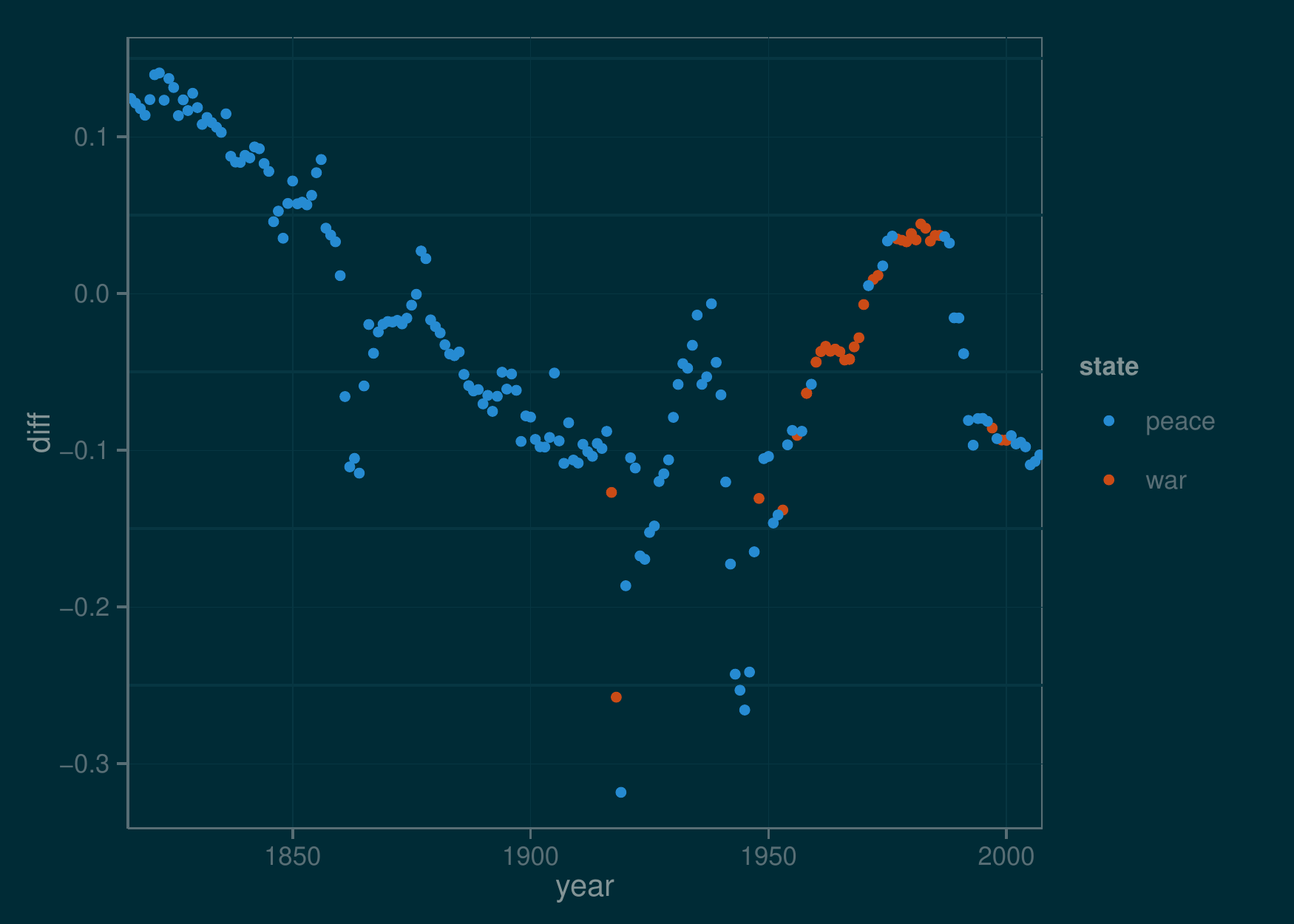}
                                  \caption{USA--Russia}
      \end{subfigure}
                 \caption{Differences of economic states between countries--opponents}
                      \label{fig:economic}
\end{figure}

Let us visualize this significance by plotting the differences of \textit{CINC} for the four selected pairs of the countries over the whole  periods of the relations between them in war\footnote{``A war between Russia and USA'' sounds dreadfully \Radioactivity. It should be noticed, however, that the COW project treats ``war'' in a generalized way, using the term ``dispute''. This can imply, for example, show of force without direct militarized actions. Such was, for instance, Cuban Missile Crisis of 1962.} and peace (Fig. \ref{fig:economic}).

In our view, the orange points of war are concentrated in the regions of extremely steep slopes of the plotted dependencies, or when a dependency sharply changes its character. Such is the \textit{CINC}--peace/war correlation. 

Differently looks case (c). We see there pretty abrupt oscillations of differences between the economic  potentials near zero. Probably, one could say about an economic rivalry correlated with a military one as a specific point of the relations between Israel and Egypt.

\subsection{Factor of religion}

 \begin{figure}[t]
         \centering
         \begin{subfigure}[]{0.48\textwidth}
                 \includegraphics[width=1\textwidth]{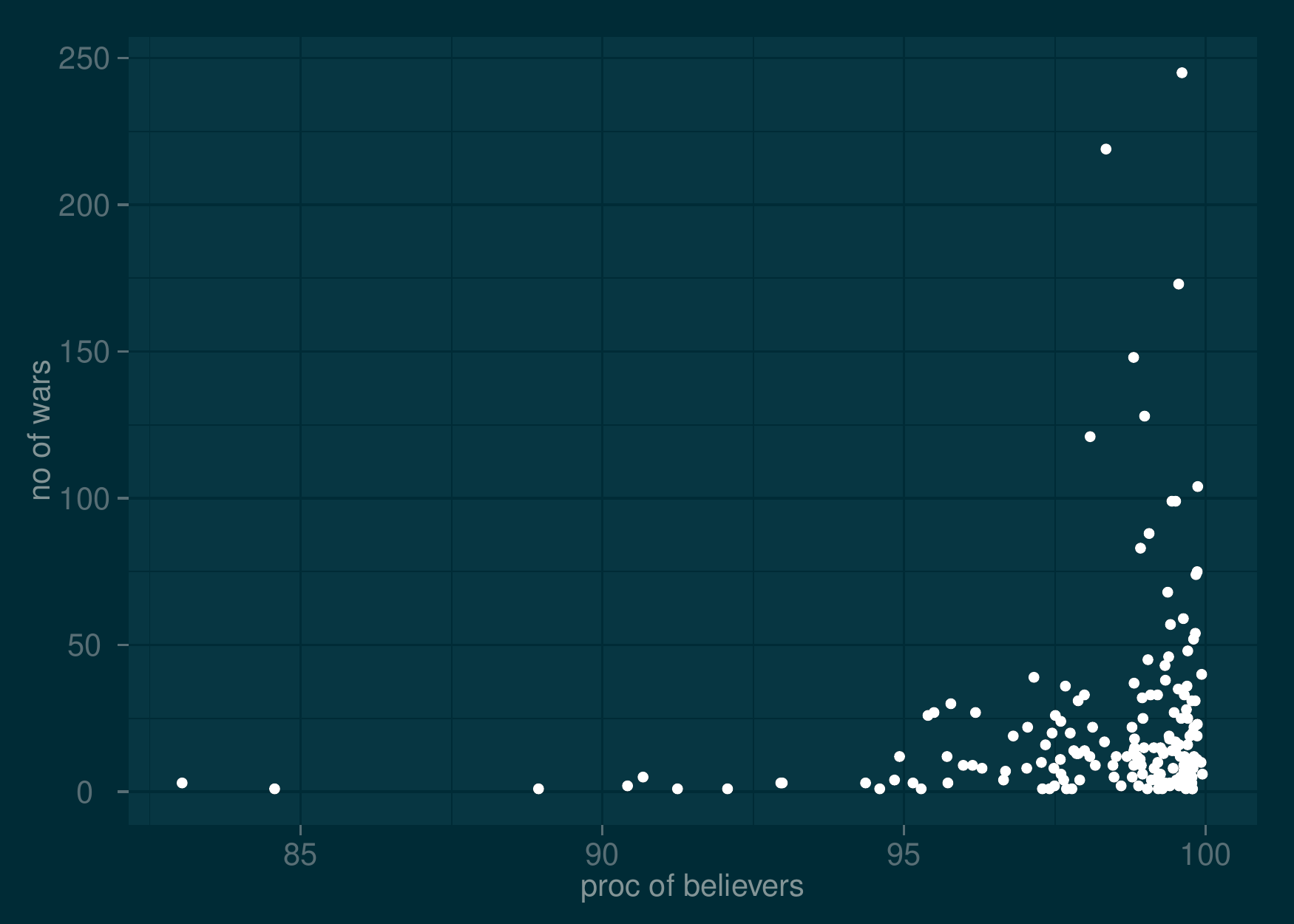}
                  \subcaption{total percentage of religious adherents}
                 \label{fig:GenBel}
         \end{subfigure}
         ~
         \begin{subfigure}[]{0.48\textwidth}
                 \includegraphics[width=1\textwidth]{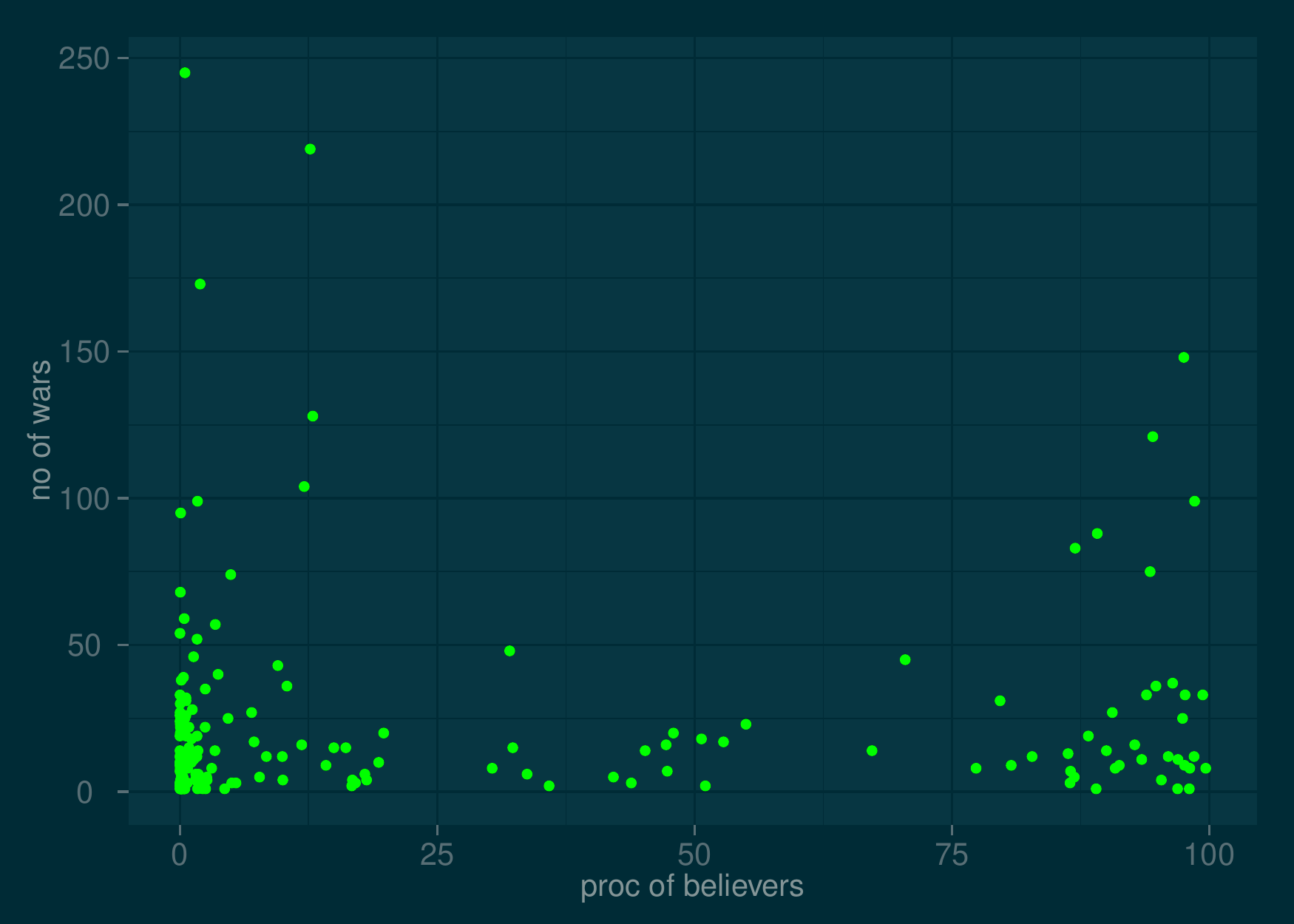}
                   \subcaption{percentage of Muslim adherents}
                 \label{fig:Islam}
         \end{subfigure}
       \caption{Countries' involvement in wars in connection with the religious factors}
            \label{fig:religion}
 \end{figure}
Nowadays  religious aspects of international tension is a matter of great public concern. That is why this factor is duly considered here. 

The Correlates of War provide a variety of data relevant to this subject as well \cite{COW_religious}. For the study we took two values: the general percentage of religious adherents in the population of a particular country, and the percentage of Islam adherents. Both values are available for every half-decade period since 1945 up to 2010.

Fig. \ref{fig:religion} shows relations between those two percentages and the number of wars the countries participated in.

 Plot (a) points out surprisingly large part of the countries which had highly religious population during last 65 years. In its turn, this part contains a few countries with extremely large number of wars in their history during this period.

Looking at plot (b) we can conclude that high involvement in wars is by no means a general feature of Islamic states. However, there are six countries of this kind with really substantial history of disputes: \textbf{\textit{Iran, Iraq, Turkey, Syria, Egypt, and Pakistan}}.

 \section{Conclusions}

 \begin{itemize}[label=\ding{43}]
 \item The tests of normality have distinguished a dozen countries with complicated history whose involvement in wars has been essentially different from the rest of the world. In our research we have been making a particular emphasis on these countries.
 \item The \textit{diplomatic activity} has turned out to significantly depend on a particular country at times of both war and peace. The tests of the whole of the countries have shown that any patterns specific for a particular state are ``averaged'' so there is no evidence of any commonly preferred ways to conduct foreign policy at times of war or peace. However, \textit{the degree of joint diplomatic activity }has proved to substantially depend upon the level of interstate tension throughout two centuries of the world history.
 \item As to \textit{the economic factor,} a strong correlation has been revealed between the difference of the opponents' \textit{Composite Indexes of National Capability} and the character of relations between pairs of the countries. Visualization of this correlations points to an onset of a dispute at the moments when the difference in the countries' economic health rapidly changes.
 \item Finally, the \textit{religious factor} has been shown to significantly correlate with the war/peace conditions. It turns out that countries with higher percentage of religious adherents have been more involved in wars during the last 65 years. As to \textit{the Islamic factor}, it looks like it hardly affects military activity greatly \textit{per se}. High involvement in wars of six large Islamic countries is evidently caused by the combination of their unique politics, economics, and culture.
 \end{itemize}
 
 No doubt, the great number and variety of \href{http://www.correlatesofwar.org/}{Correlates of War} data contain more features, relations, and causalities. So the continuation of this research may follow in some future.

\bibliographystyle{plain}
\bibliography{MackarovSolution}
\end{document}